\documentclass{PoS}

\usepackage{amsmath}
\usepackage{amsfonts}
\usepackage{float}
\usepackage{booktabs}

\usepackage{subfig}

\include{subfig}

\newcommand{\imineq}[2]{\vcenter{\hbox{\includegraphics[height=#2ex]{#1}}}}

\title{NLO mixed QCD-EW corrections to Higgs \\gluon fusion}

\ShortTitle{NLO QCD-EW $gg \to H$}

\author{\speaker{Marco Bonetti}\\
        Institute for Theoretical Particle Physics, KIT, Karlsruhe, Germany\\
        E-mail: \email{marco.bonetti@kit.edu}}

\abstract{The study of the Higgs boson properties is one of the main tasks of contemporary high-energy physics. Among Higgs properties, its interaction with gluons is interesting since it can be facilitated by yet unknown elementary particles. One of the major sources of uncertainty in the theoretical description of $ggH$ coupling originates from mixed QCD-electroweak contributions. The NLO QCD corrections to these contributions were evaluated in the approximation where electroweak boson masses were considered to be significantly larger than the mass of the Higgs boson and it is desirable to compute these corrections for physical masses of the gauge bosons and the Higgs boson. We present a major step towards this goal and describe first the analytic evaluation of NLO mixed QCD-EW three-loop virtual corrections to $gg \to H$, and then their implementation in the evaluation of the total cross section for $gg \to H$ in the soft-gluon approximation for real corrections.
\\\\\\
\hspace*{\fill}TTP18-029}

\FullConference{Loops and Legs in Quantum Field Theory (LL2018)\\
		29 April 2018 - 04 May 2018\\
		St. Goar, Germany}

\begin{document}


\section{Motivations}
The Standard Model has proven to be a solid framework to describe elementary particles, although in recent times more and more evidences demand an extension of the current theory. Together with a lack of direct detection of new particles at colliders, the current situation requires an increase in the precision of measurements and theoretical calculations to investigate indirect signatures of beyond the Standard Model physics.

The Higgs boson represents a good candidate to pursue this research project: its properties and couplings are still under investigation, and on general grounds we expect a change in the $ggH$ coupling due to new physics at $\mathcal{O}(1\,\textup{TeV})$ to be around $5\%$.

This sets the goal for theoretical uncertainties to be below $\mathcal{O}(1\%)$. Considering gluon fusion, the main channel of production of Higgs bosons at the LHC, the theoretical uncertainty (see \cite{Anastasiou:2016cez,Patrignani:2016xqp,Mistlberger:2018etf}) at present originates from $\mathcal{O}(2\%)$ residual scale uncertainty in pure QCD contributions, $\mathcal{O}(1\%)$ uncertainty caused by unknown mass effects of $b$ and $c$ quarks in higher order QCD corrections, and $\mathcal{O}(1\%)$ uncertainty in QCD-EW contributions. The case of QCD-EW corrections is peculiar. Indeed, although the LO ($\alpha_{~}^2 \alpha_S$) contribution is know for arbitrary values of $m_H$ and $m_{W,Z}$ \cite{Aglietti:2004nj,Aglietti:2006yd}, QCD corrections ($\alpha_{~}^2 \alpha_S^2$) to this kind of diagrams are known only using different approximations: either in the limit $m_H \ll m_{W,Z}$  \cite{Anastasiou:2008tj} or using factorization approaches \cite{Actis:2008ug,Actis:2008ts}. Corrections to the fermionic channel has been computed in \cite{Degrassi:2004mx}. To decrease the corresponding theoretical uncertainty it is therefore necessary to evaluate the NLO contributions to mixed QCD-EW corrections to Higgs production in gluon fusion for physical values of $m_H$ and $m_{W,Z}$ and full QCD structure \cite{Bonetti:2016brm,Bonetti:2017ovy,Bonetti:2018ukf}.


\section{Amplitude}
\begin{figure}
	\centering
	\subfloat[][LO.]
	{\includegraphics[width=0.47\textwidth]{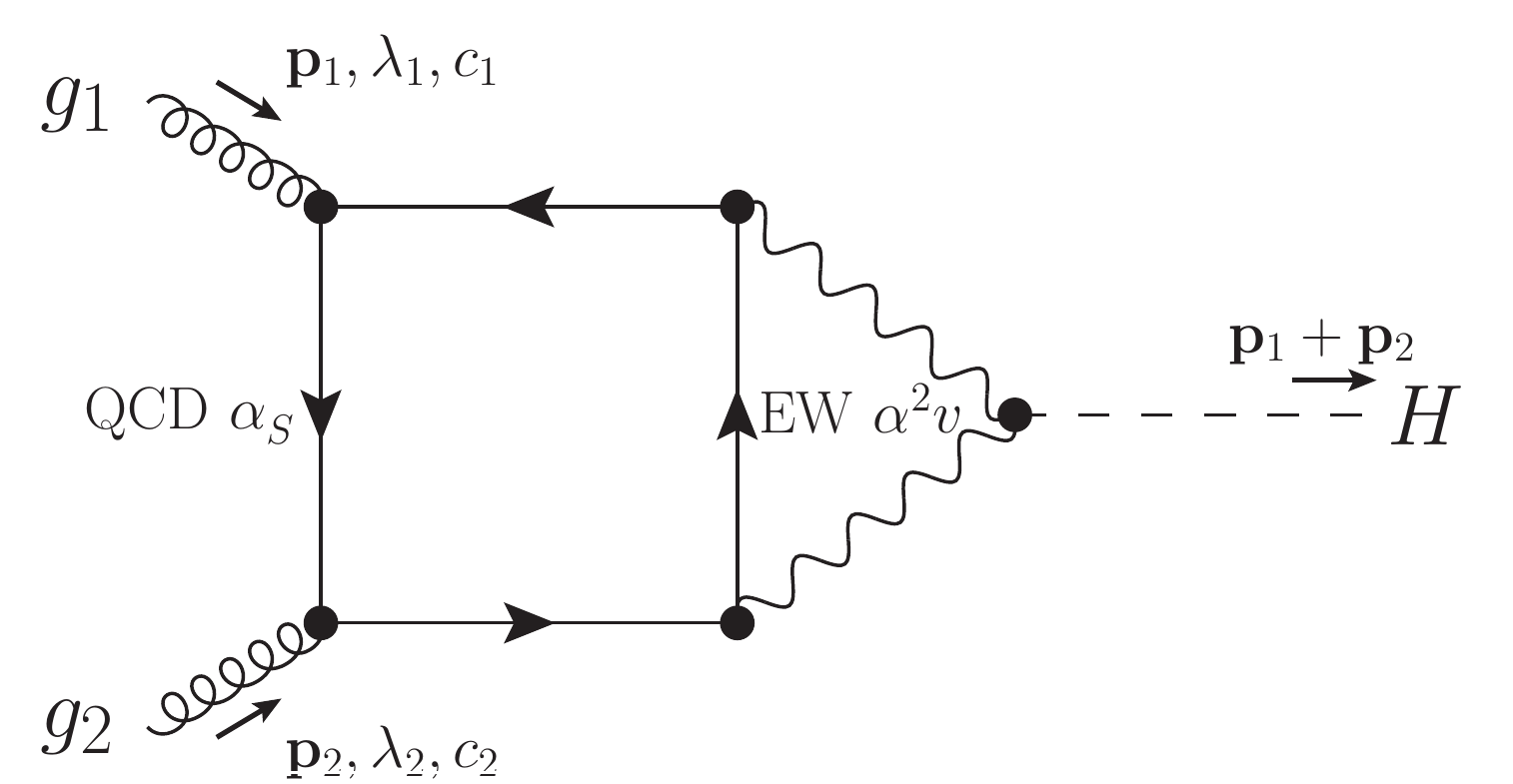}}
	\qquad
	\subfloat[][Virtual NLO.]
	{\includegraphics[width=0.47\textwidth]{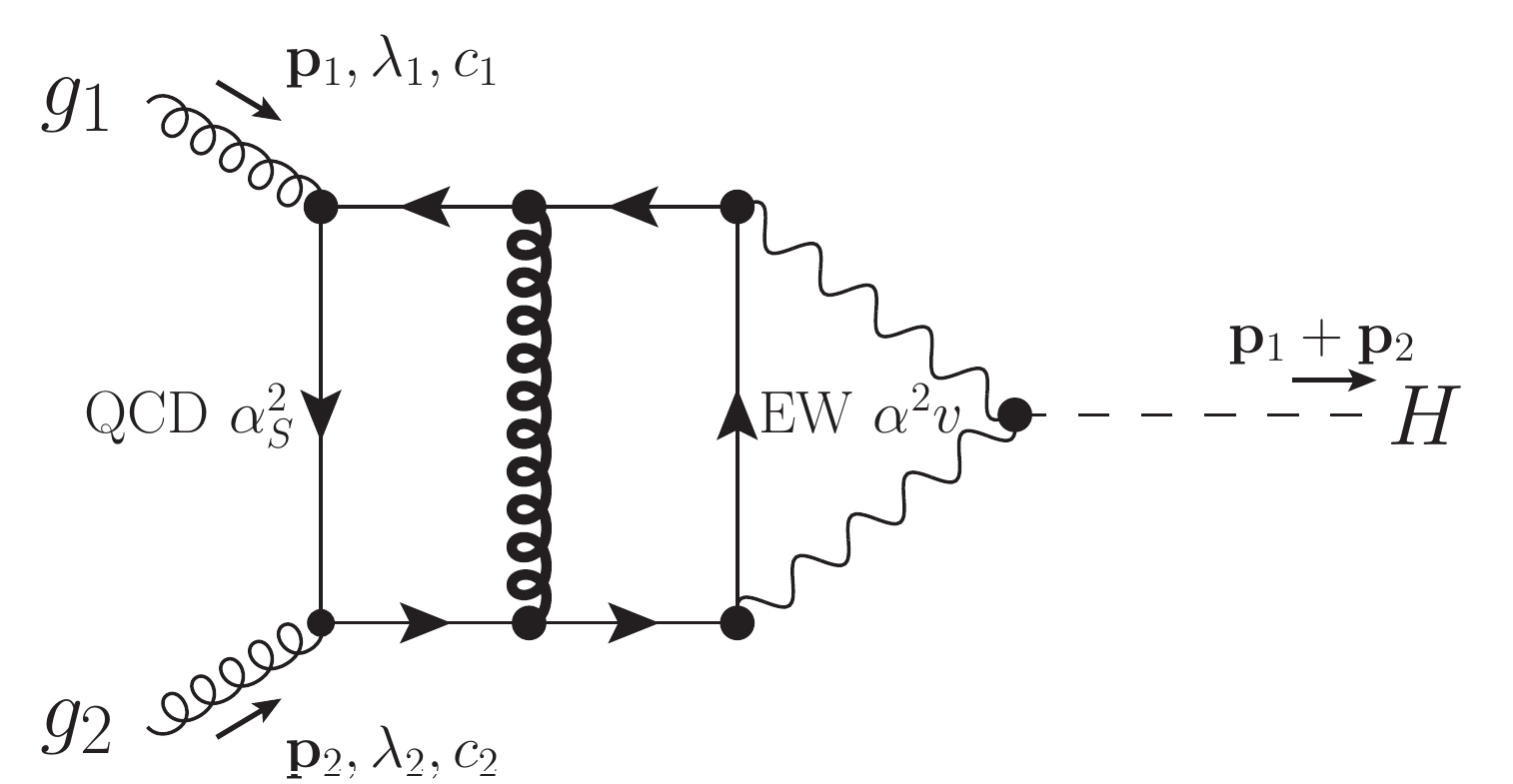}}
	\caption{Sketch of Feynman diagrams for mixed QCD-EW contributions to $gg \to H$.}
	\label{QCD-EW}
\end{figure}

To compute the NLO cross-section we start considering QCD virtual corrections to the LO QCD-EW contributions to $gg \to H$, as shown in Fig.~\ref{QCD-EW}.

In variance with the pure QCD case, where the main contribution is given by the top quark coupling to the Higgs boson, in this class of QCD-EW corrections the Higgs boson always couples to EW vector bosons, which subsequently couple to the QCD part through a fermionic loop. The first effect of this modified coupling is that the top quark contribution is suppressed with respect to the contribution of light quarks \cite{Degrassi:2004mx}, since the enhancing factor given by the Yukawa coupling is not present. We can therefore consider light quarks only (the first two generations for the diagrams containing $W$ bosons, all quarks but the top in case of $Z$ bosons), and work in the approximation of massless quarks. Moreover, since $W$ and $Z$ bosons never appear together in the same diagram, each single diagram depends on two parameters only: the square of the mass of the EW vector boson $m_V^2$ and the square of the mass of the Higgs boson $s=m_H^2=(p_1+p_2)^2$.

There are 47 Feynman diagrams both for $W$ and $Z$ bosons coupling to $H$. The amplitude consists of a single form factor
\begin{equation}
	\mathcal{M}^{c_1 c_2}_{\lambda_1 \lambda_2} = \delta^{c_1 c_2} \epsilon_{\lambda_1}(\mathbf{p}_1) \cdot \epsilon_{\lambda_2}(\mathbf{p}_2) \mathcal{F}(s,m_W,m_Z).
\end{equation}
In particular
\begin{equation}
	\label{eqa}
	\mathcal{F}(s,m_W,m_Z) = -\mathrm{i}\frac{\alpha_{~}^2 \alpha_S(\mu) v}{64 \mathrm{\pi} \sin^4\theta_W} \sum_{V=W,Z} C_V A(m_V^2/s,\mu^2/s),
\end{equation}
where
\begin{equation}
	C_W = 4,	\qquad\qquad 	C_Z = \frac{2}{\cos^4\theta_W}\left( \frac{5}{4}-\frac{7}{3}\sin^2\theta_W+\frac{22}{9}\sin^4\theta_W \right).
\end{equation}
The quantity $A$ in Eq.~(\ref{eqa}) can be naturally expanded in powers of $\alpha_S$. This gives
\begin{equation}
	A(m^2/s,\mu^2/s) = A_\textup{2L}(m^2/s) + \frac{\alpha_S(\mu)}{2\mathrm{\pi}} A_\textup{3L}(m^2/s,\mu^2/s) + \mathcal{O}\left( \alpha_S^2 \right).
\end{equation}

The three-loop contribution $A_\textup{3L}(m^2/s,\mu^2/s)$ can be written as a linear combination of 95 three-loop scalar \emph{master integrals} $\mathcal{I}(s,m,\epsilon)$, where $\epsilon=(4-D)/2$ is the dimensional regularization parameter.

\section{Differential equations and uniformly transcendental functions}
It is useful to change variables from $(s,m^2)$ to $(s,y)$, where $y$ is defined as
\begin{equation}
	y = \frac{\sqrt{1-4m^2/s}-1}{\sqrt{1-4m^2/s}+1}.
\end{equation}
With these new variables, the $s$-dependence of the master integrals can be determined simply by dimensional analysis of the integrals. This gives
\begin{equation}
	\mathcal{I}_n(s,y,\epsilon) = (-s-\mathrm{i}0)^{a_n-3\epsilon}\mathcal{J}_n(y,\epsilon).
\end{equation}
We determine the functions $\mathcal{J}_n(y,\epsilon)$ using differential equations. We derive them by differentiating the master integrals with respect to $y$ and we include in cascade DEs for subtopologies appearing in the r.h.s. until we obtain a closed system of first-order differential equations. We then proceed to solve this system of equations.

As shown in Ref.~\cite{Henn:2013pwa,Argeri:2014qva}, in many cases of interest it is possible to express the master integrals using so-called \emph{uniformly transcendental functions}, i.e. functions that admit series expansion in $\epsilon$ with coefficient of the $\epsilon^n$ term having weight $n$. A function $F_n(y)$ has \emph{weight} $n$ if it can be written as $n$ nested integrations over $\mathrm{d}\log R_n(\xi)$, where $R_n(\xi)$ is a rational function in $\xi$
\begin{equation}
	F_n(y) = \int_0^y \dots \int_0^{\xi_{n-1}} \,\mathrm{d}\log R_n(\xi_n) \dots \mathrm{d}\log R_1(\xi_1)	\qquad \Rightarrow \qquad 	w(F_n) = n.
\end{equation}
This definition of weight admits an unambiguous extension to constants: weight $n$ functions evaluated in rational points give weight $n$ constants. Moreover, a function which is a product of two functions with weights $n_1$ and $n_2$ has weight $n_1+n_2$.

Uniformly transcendental functions satisfy a characteristic Cauchy problem \cite{Remiddi:1999ew,Henn:2013pwa,Lee:2014ioa}. Their differential equations can be written in the so-called \emph{canonical fuchsian} form
\begin{equation}
	\frac{\mathrm{d}}{\mathrm{d}y}\mathbf{F}(y,\epsilon) = \epsilon \sum_{a=1}^A B_a \frac{\mathrm{d}\log R_a(y)}{\mathrm{d}y}\mathbf{F}(y,\epsilon),
\end{equation}
where the $\epsilon$ dependence is completely factorized, $B_a$ are matrices which elements are rational numbers, and kinematic variables are present only in simple poles coming from the $\mathrm{d}\log R(y)/\mathrm{d}y$ structures, where $R(y)$ is a rational function of $y$.

Furthermore, it is much easier to fix the integration constants for uniformly transcendental functions: by comparing the solution of the differential equations with a boundary value at a rational point $y_0$ it is possible to express the integration constants in terms of simple rational combinations of a small set of constant with weight given by the order in the $\epsilon$ expansion. In other words, the boundary value $L(y,\epsilon)$ in $y \to y_0$ is a uniformly transcendental sum of constants.

In order to exploit these useful properties of uniformly transcendental functions, we proceed to tune our system of differential equations and boundary conditions in order to obtain a canonical system of equations and uniformly transcendental boundary functions. In our case we explicitly cast the system of differential equations into a canonical fuchsian form, while we investigate directly the boundary conditions only for the simplest subtopologies, common to all the master integrals. This procedure has proven to be sufficient in our case to find uniformly transcendental expressions for all our master integrals.

Starting from the Cauchy problem for uniformly transcendental functions, its solution can be written in terms of a Dyson series in $\epsilon$ thanks to the canonical structure of the DEs \cite{Remiddi:1999ew,Henn:2013pwa,Argeri:2014qva,Lee:2014ioa}
\begin{multline}
	\mathbf{F}(y,\epsilon) = \mathcal{P}_y\mathrm{e}^{\epsilon \int A(\xi)\,\mathrm{d}\xi}\mathbf{F}_0(\epsilon) = F_0^{(0)} + \left[ \int_y A(\xi_1) \mathbf{F}_0^{(0)}\,\mathrm{d}\xi_1 + \mathbf{F}_0^{(1)} \right] + \\
	+ \left[ \int_y A(\xi_1) \int_{\xi_1} A(\xi_2) \mathbf{F}_0^{(0)}\,\mathrm{d}\xi_2\mathrm{d}\xi_1 + \int_y A(\xi_1) \mathbf{F}_0^{(1)}\,\mathrm{d}\xi_1 + \mathbf{F}_0^{(2)} \right] + \mathcal{O}\left( \epsilon^3 \right)
\end{multline}
where $A(\xi) = \sum_{a=1}^A B_a \mathrm{d}\log R_a(\xi)/\mathrm{d}\xi$.

Nested integrations are naturally expressed in terms of \emph{Goncharov Polylogarithms} \cite{Goncharov}
\begin{equation}
	G(a_n,a_{n-1},\dots,a_1;y) := \int_0^y \frac{1}{\xi-a_n} G(a_{n-1},\dots,a_1;\xi) \,\mathrm{d}\xi
\end{equation}
with $G(z) := 1$ and $G(0_n,\dots,0_1;y) := \log^n y / n!$.

The integration constants $F_0^{(n)}$ are, order by order in $\epsilon$, simple rational linear combinations of fixed weight constants. The constants that appear in our calculation are enlisted in Table~\ref{table1}.

\begin{table}
\centering
\begin{tabular}{ccc}
\toprule
Weight	&\multicolumn{2}{c}{Values}				\\
\midrule
0	&\multicolumn{2}{c}{$1$}				\\
1	&\multicolumn{2}{c}{$\emptyset$}			\\
2	&\multicolumn{2}{c}{$\mathrm{\pi}^2$}			\\
3	&\multicolumn{2}{c}{$\zeta(3)$}				\\
4	&\multicolumn{2}{c}{$\mathrm{\pi}^4$}			\\
5	&$\mathrm{\pi}^2 \zeta(3)$		&$\zeta(5)$	\\
6	&$\mathrm{\pi}^6$			&$\zeta^2(3)$	\\
\bottomrule
\end{tabular}
\caption{Values appearing in the constant terms at each weight.}
\label{table1}
\end{table}

Given their small number, it is possible to fit them using numerical evaluation with high number of digits at the boundary and to compare the result to the expected boundary value $\mathbf{L}(y,\epsilon)$
\begin{equation}
	\lim_{y \to 1} \left[ \mathbf{F}(y,\epsilon)-\mathbf{L}(y,\epsilon) \right].
\end{equation}

\section{Calculation of the virtual NLO QCD-EW contributions}
The system of DEs for the 95 MIs of $A_\textup{3L}(m^2/s,\mu^2/s)$ in canonical fuchsian form reads
\begin{multline}
	\mathrm{d}\mathbf{F}(y,\epsilon) = \epsilon \left[ B_+ \, \mathrm{d}\log(1-y) + B_r \, \mathrm{d}\log(y^2-y+1) + B_- \, \mathrm{d}\log(y+1) + B_0 \, \mathrm{d}\log y \right] \mathbf{F}(y,\epsilon).
\end{multline}
The $\mathrm{d}\log$ structures appearing in the the system (the same appearing in the differential equations for $A_\textup{2L}(m^2/s)$) can be related to the cuts on the diagrams: $(1-y)$ corresponds to the cut of all massless lines, $(y^2-y+1)$ to the cut of just one massive line, $(y+1)$ to the cut of two massive lines, and $y$ to the residue at $s \to \infty$.

To fix the integration constants we choose to compare the results from the differential equations with an independent evaluation at the point $y \to 1$. This particular point allows for a fast numerical evaluation of GPLs at high precision, as well as for a clear physical interpretation of the result, given by the fact that $y \to 1$ corresponds to the limit $m^2 \gg s$. We can therefore compute the boundary values of the integrals for the matching by performing a \emph{large-mass expansion} of our functions \cite{Smirnov:2002pj}.

We compute the large-mass expansion for the master integrals in the following way. MIs depend on two different scales: the external momenta $p_{1,2}$ and the internal mass $m$. The large-mass expansion corresponds to the \emph{mathematical limit} $p_1 \sim p_2 \sim \sqrt{s} \ll m$, where configurations of loop momenta $\{k\}_I$ that can produce a non-vanishing contribution correspond to internal momenta either scaling as $\sqrt{s}$ or $m$. These contributions must also satisfy a ``large-momentum conservation law'', which states that large momentum cannot be created, destroyed or provided by external legs. This forces lines carrying large momentum to be internal ones and to be arranged in a closed circuit. After Taylor expanding the integrand in all small parameters up to the required order, we sum over all non-vanishing integrals to obtain the large-mass expansion. Notice that this procedure allows for a diagrammatic approach, giving us a collection of tadpoles, massless bubbles and massless triangles, all available in the literature.

As an example, consider the integral
\begin{equation}
	\imineq{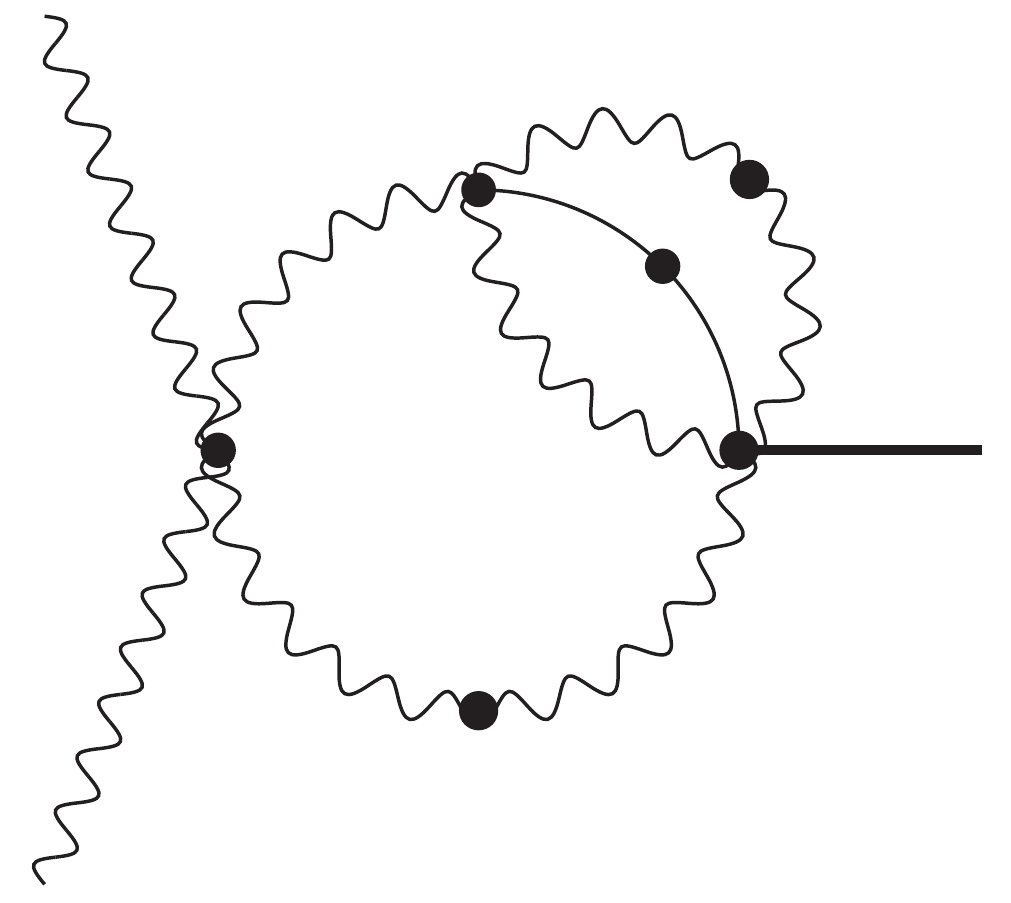}{12}.
\end{equation}
Its large-mass expansion consists of two terms (thick lines indicate large momentum flow)
\begin{align}
	\imineq{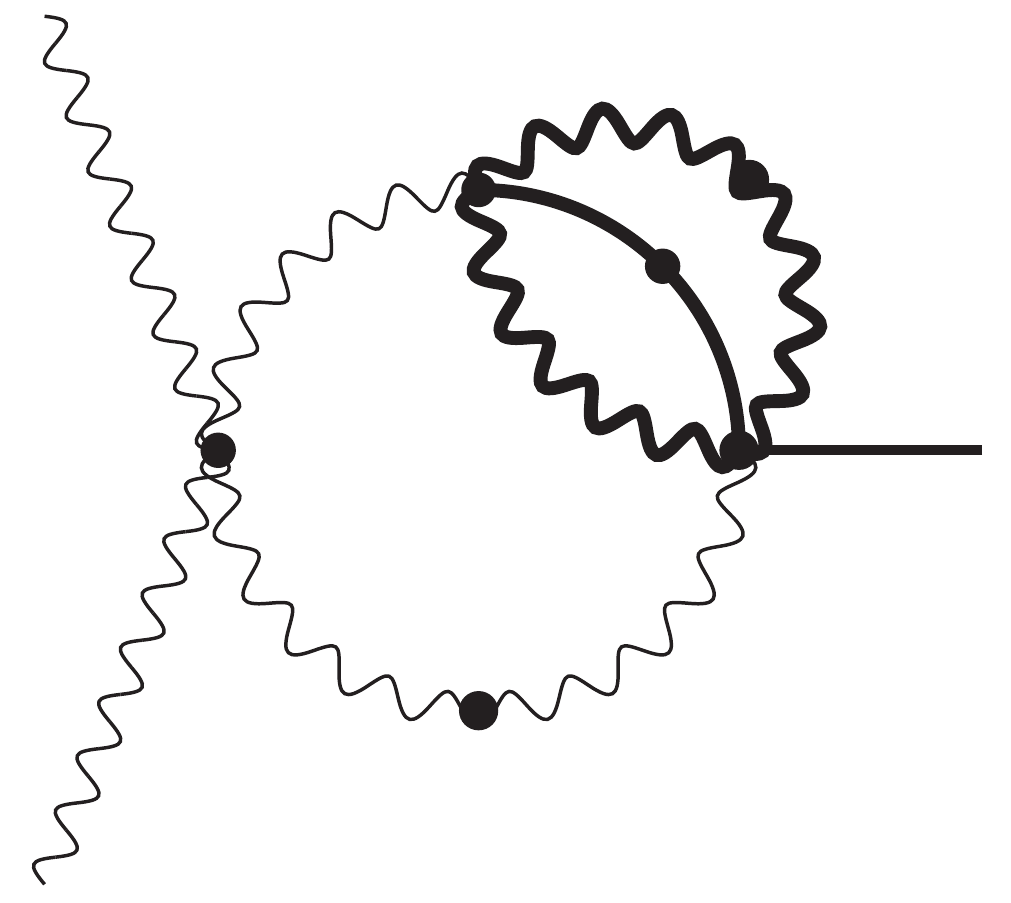}{10} \quad\to\quad& \imineq{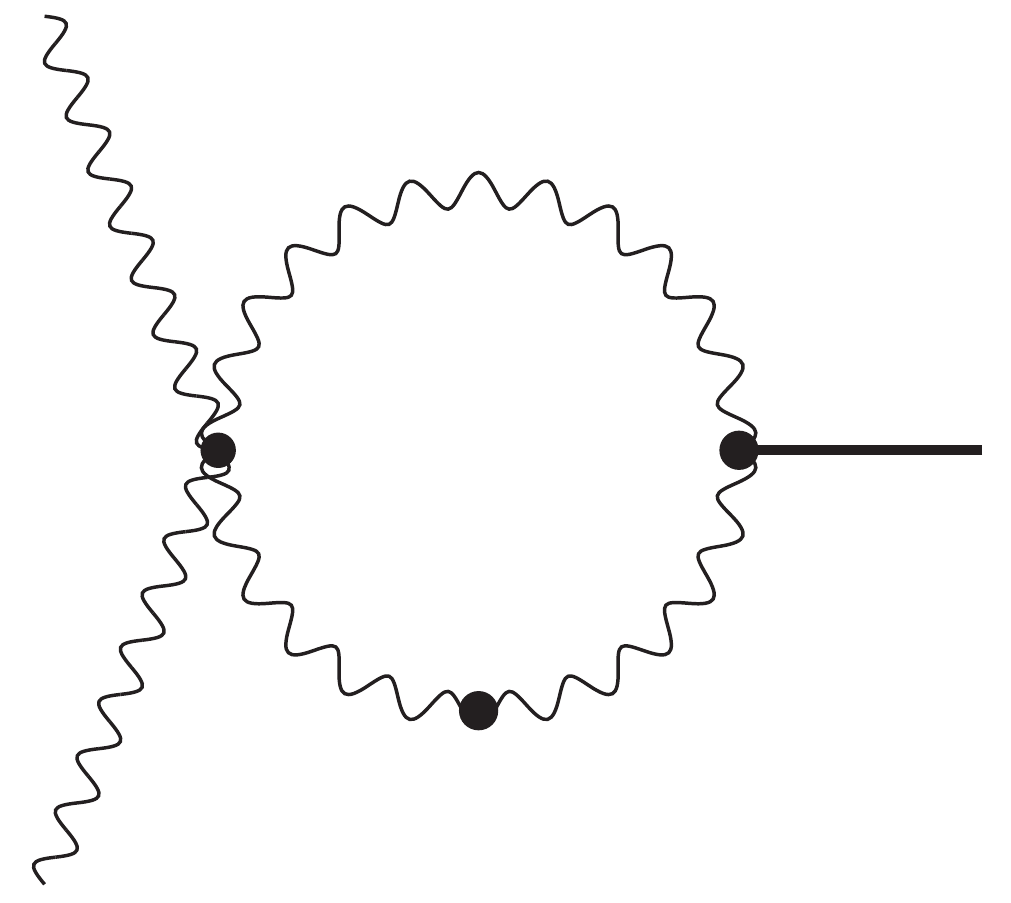}{10} \quad\times\quad \imineq{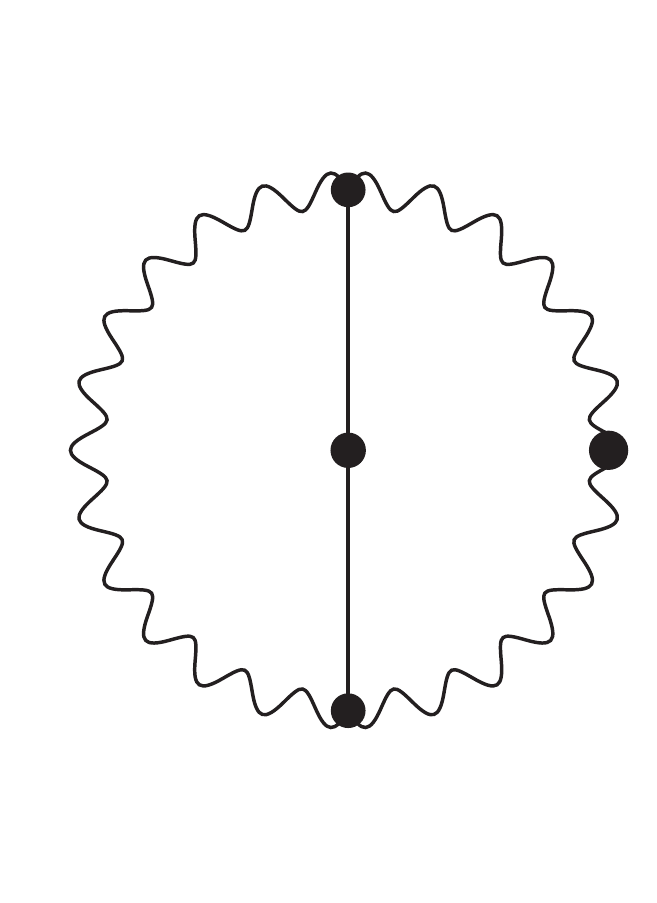}{10},	\\
	\imineq{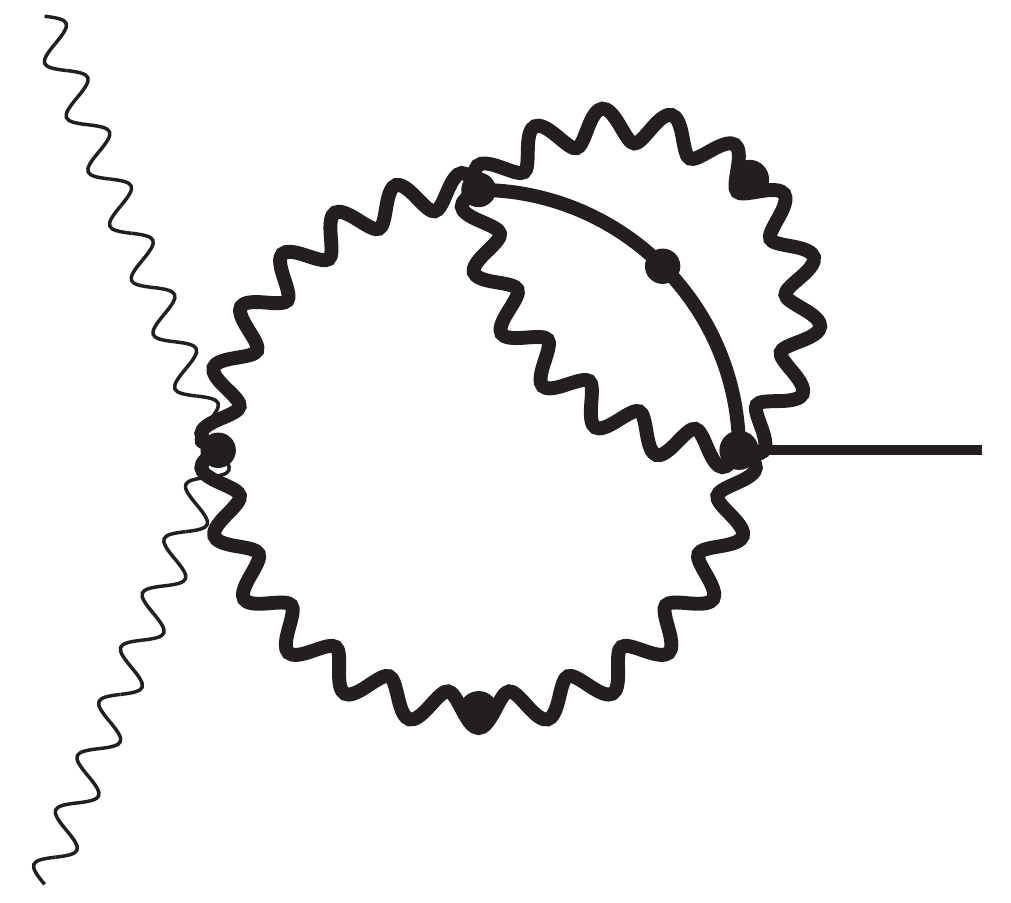}{10} \quad\to\quad& \imineq{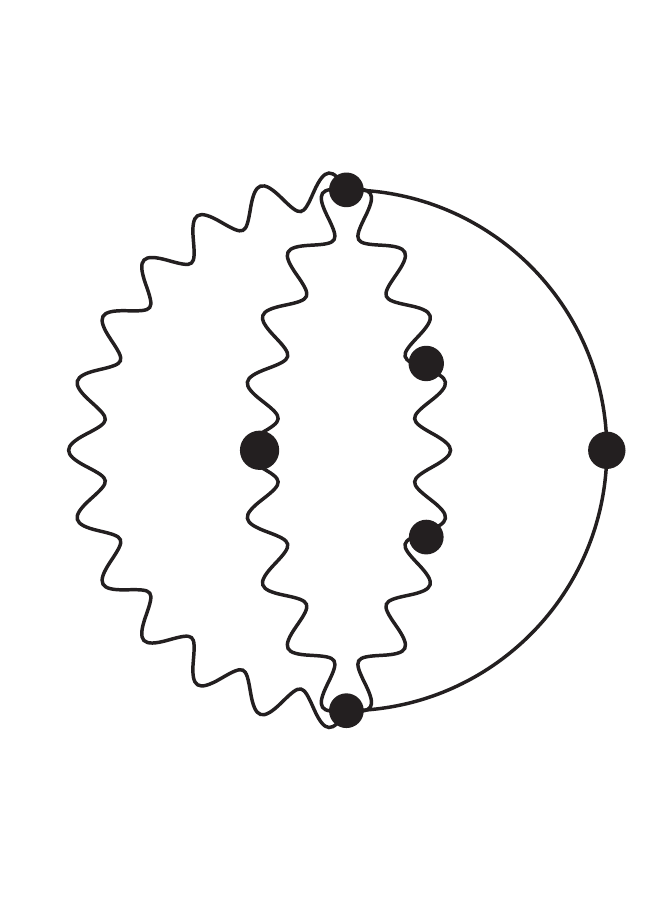}{10} \quad+\quad s\frac{2(1+\epsilon)}{2-\epsilon}\,\imineq{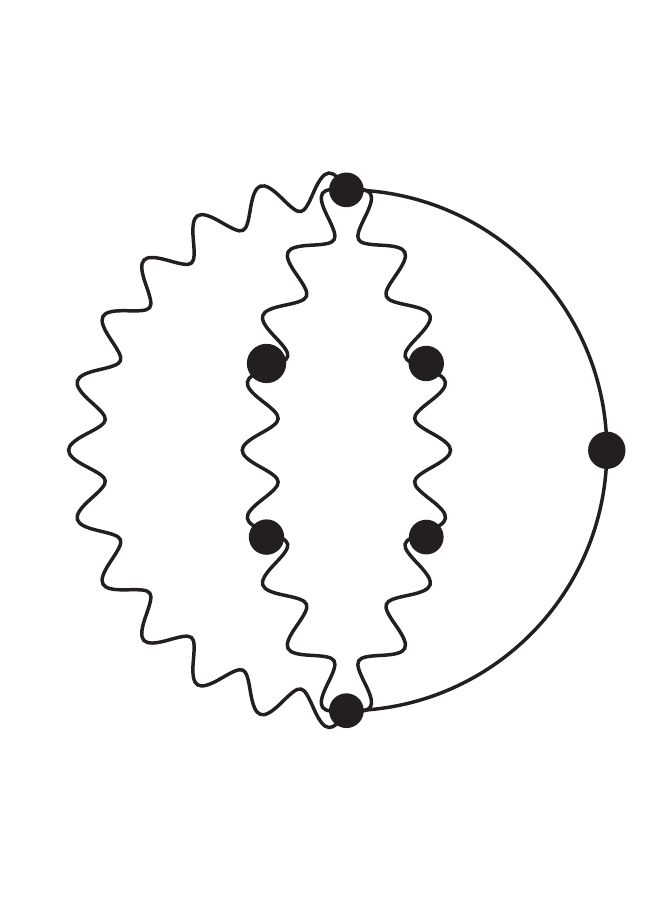}{10} \quad+\quad \mathcal{O}\left(\frac{(-s)^2}{(M^2)^{4}}\right).
\end{align}

All the MIs have been numerically checked against \texttt{SecDec} \cite{Borowka:2015mxa} and \texttt{pySecDec} \cite{Borowka:2017idc} for multiple values of $y$. Agreement was found in all cases.

\section{NLO cross-section}
As expected, virtual NLO contribution shows both UV and IR divergencies. The UV divergencies are fixed by renormalization of $\alpha_S$ (since we are considering QCD corrections to LO amplitude). The IR divergent part has to cancel against real corrections. Its form is given by Catani's formula \cite{Catani:1998bh}
\begin{gather}
	A_\textup{3L} = \mathbf{I}_g^{(1)} A_\textup{2L} + A_\textup{3L}^\textup{fin},	\\
	\mathbf{I}^{(1)}=\left(\frac{-s-i0}{\mu^2}\right)^{-\epsilon}
\frac{\mathrm{e}^{\epsilon\gamma_E}}{\Gamma(1-\epsilon)}
\left[-\frac{C_A}{\epsilon^2}-\frac{\beta_0}{\epsilon}\right].
\end{gather}
where $A_\textup{2L}$ up to $\mathcal{O}\left(\epsilon^2\right)$ was computed in \cite{Bonetti:2016brm}.

\begin{figure}
	\centering
	\subfloat[][Cut for $s=0$ at LO.]
	{\includegraphics[width=0.40\textwidth]{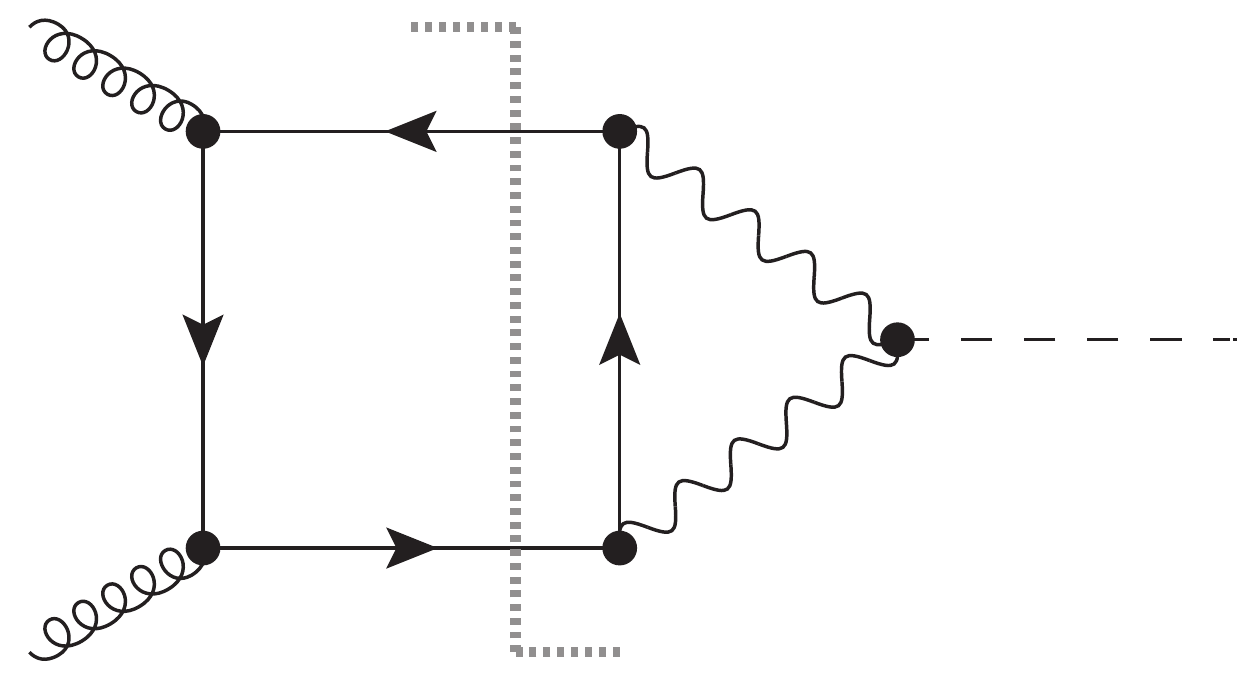}}
	\qquad
	\subfloat[][Non-zero contribution for $s > 0$ at NLO.]
	{\includegraphics[width=0.40\textwidth]{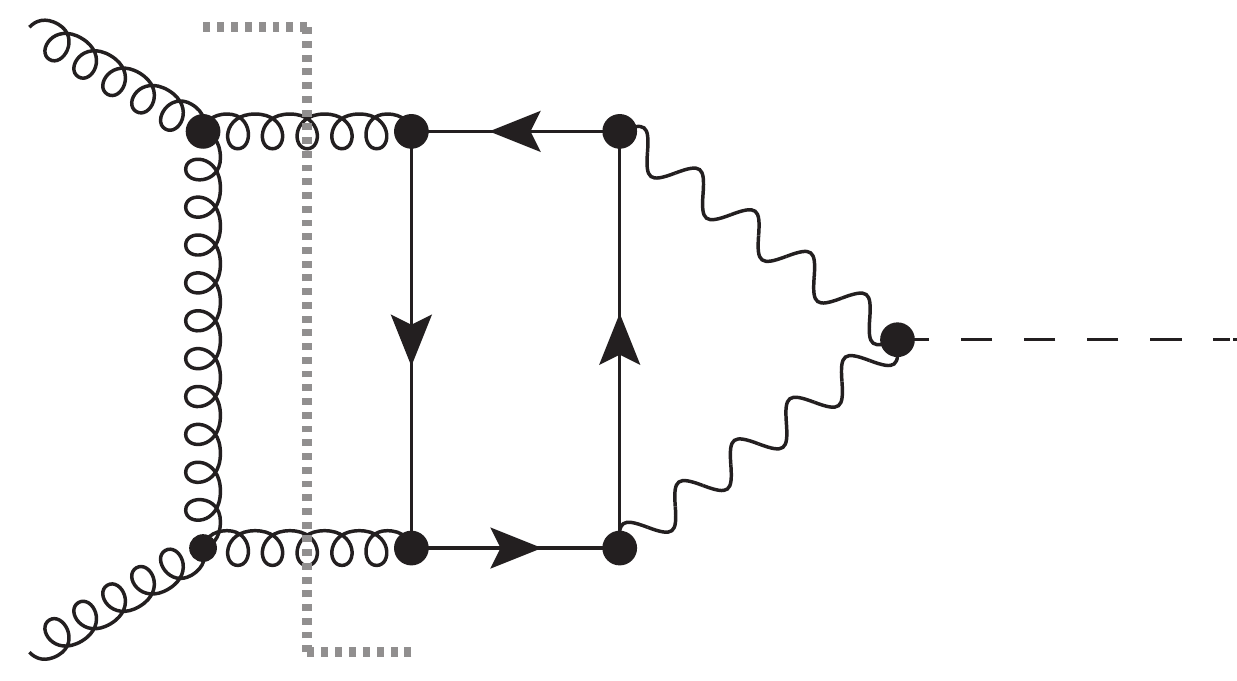}}
	\caption{Contributions from diagrams like the one on the left vanish in the sum at LO. Diagrams on the right provide a non-zero contribution in the sum at NLO.}
	\label{cuts}
\end{figure}

Taking $\sqrt{s}=\mu=m_H=125.09\,\textup{GeV}$, $m_W=80.385\,\textup{GeV}$, $m_Z=91.1876\,\textup{GeV}$, $N_C=3$ and $N_f=5$ we obtain
\begin{equation}
	\label{eq1}
	\begin{array}{lcll}
		A_\text{LO}(m_Z^2/m_H^2,1)		&=	&-6.880846	&- \mathrm{i}\, 0.5784119\,, \\
		A_\text{LO}(m_W^2/m_H^2,1)		&=	&-10.71693	&- \mathrm{i}\, 2.302953\,, \\
		A^\text{fin}_\text{NLO}(m_Z^2/m_H^2,1)	&=	&-2.975801	&- \mathrm{i}\, 41.19509\,, \\
		A^\text{fin}_\text{NLO}(m_W^2/m_H^2,1)	&=	&-11.31557	&- \mathrm{i}\, 54.02989\,.
	\end{array}
\end{equation} 
It is possible to see that the difference between the imaginary parts of the amplitude is much bigger than the difference in the real parts. This behavior can be understood since imaginary parts of the amplitude are related to the possibility of producing on-shell intermediate particles. At the level of single diagrams, crossing both $s=0$ and $s=m_V^2$ generates imaginary parts ($s=0$ for on-shell massless fermions and $s=m_V^2$ for one on-shell massive vector boson). At the level of the amplitude (summing all diagrams) the contributions at $s=0$ for the 2-loop case vanish, since the only possible cut generates processes of the form $gg \to q \bar{q} \quad|\quad q \bar{q} \to H$, and the Higgs boson cannot couple to massless fermions, even at the loop level. This is not the case at three loops, since other cuts can give a non vanishing contribution crossing $s=0$, as depicted in Fig.~\ref{cuts}.

To obtain a value for the NLO QCD-EW contributions to the physical cross-section real corrections are required. A first result can be obtained considering that for Higgs gluon fusion in the pure QCD case it has been observed that the gluonic PDFs suppress real corrections with a highly energetic extra gluon. Assuming this property to be valid also for QCD corrections to QCD-EW processes, we can employ a soft limit for the extra gluon, which allows us to factorize the real emission from the partonic squared amplitude
\begin{equation}
	\lim_{E_4 \to 0} \left| A_\textup{NLO}^\textup{real} \right|^2 = \frac{\alpha_S^{~}}{4\mathrm{\pi}} N_C \frac{2 p_1 \cdot p_2}{p_1 \cdot p_4 p_2 \cdot p_4} \left| A_\textup{LO}^{~} \right|^2 + \mathcal{O}\left( p_4^{-1} \right),
\end{equation}
where $p_4$ and $E_4$ are the 4-momentum and the energy, respectively, of the extra gluon.

At the level of the hadronic cross-section we have \cite{Catani:2001ic,deFlorian:2012za,Bonvini:2013jha}
\begin{equation}
	\label{cs}
	\sigma = \int_0^1 \int_0^1 f(x_1,\mu) f(x_2,\mu) \sigma_\textup{LO} z G(z,\mu,\alpha_S^{~}) \,\mathrm{d} x_2 \mathrm{d} x_1,
\end{equation}
where $z:=m_H^2/(S_\textup{h} x_1 x_2)$ is the energy of the core process $gg \to H$, and
\begin{gather}
	G = \delta(1-z) + \frac{\alpha_S}{2\mathrm{\pi}}\left [8 N_C \left( \mathcal{D}_1 + \frac{\mathcal{D}_0}{2} \log \frac{m_H^2}{\mu^2} \right)
+ \left ( \frac{2 \pi^2}{3} N_C + \frac{\sigma_\textup{NLO}^\textup{fin}}{\sigma_\textup{LO}} \right ) \delta(1-z) \right ],	\\
	\mathcal{D}_0 := \left[ \frac{\log(1-z)}{1-z} \right]_+,	\\
	\mathcal{D}_1 := \left[\frac{\log(1-z)}{1-z}\right]_+ + (2-3z + 2 z^2) \frac{\log[ (1-z)/\sqrt{z}]}{1-z} - \frac{\log(1-z)}{1-z},
\end{gather}
with $\sigma_\textup{NLO}^\textup{fin}$ being the cross-section contribution coming from the NLO finite remainder from Eq.~(\ref{eq1}).

The numerical evaluation of the hadronic cross-section in Eq.~(\ref{cs}) using NNPDF30 for PDFs and the running of $\alpha_S$ gives
\begin{equation}
	\begin{array}{lclcccrl}
		\sigma_\textup{LO}^\textup{QCD} = 20.6\,\textup{pb},		&~~~~~~~	&\sigma_\textup{LO}^\textup{QCD-EW} = 21.7\,\textup{pb}		&~~~	&\Rightarrow	&~~~	&+ 5.3 \%	&\textup{~at~LO}\\
		\sigma_\textup{NLO}^\textup{QCD} = 32.7\,\textup{pb},		&~~~~~~~	&\sigma_\textup{NLO}^\textup{QCD-EW} = 34.4\,\textup{pb}	&~~~	&\Rightarrow	&~~~ &+5.2 \%	&\textup{~at~NLO}
  	\end{array}
\end{equation}
showing that the enhancement given by QCD corrections is similar between pure QCD and QCD-EW $gg \to H$.

\section{Conclusions}
We evaluated the NLO mixed QCD-EW corrections to $gg \to H$, employing the soft-gluon limit for real emissions. This provides a modification of the cross-section from LO to NLO of $+5.2\%$, in line with the corresponding enhancement in the pure QCD case.

The error previously associated with the QCD-EW contributions, coming from the discrepancy in the results obtained in complete factorization and in EFT for $m_{W,Z} \to +\infty$, has been removed by the present calculation.

The next necessary step towards a full result is given by the exact evaluation of the real corrections, featuring $gg \to gH$, $q\bar{q} \to gH$, $qg \to qH$ and $\bar{q}g \to \bar{q}H$ contributions. This is an interesting and challenging task both for physics and mathematics.

\bibliographystyle{BiBTeX/BIBLIOSTYLE.bst}
\bibliography{BiBTeX/biblio.bib}

\end{document}